\documentstyle[12pt]{article}
\advance\textheight by 60pt
\advance\voffset by -40pt
\advance\textwidth by 50pt
\advance\oddsidemargin by -25pt
\advance\evensidemargin by -25pt

\def\single_space{\baselineskip 12pt plus 1pt minus 1pt}
\def\one_and_a_half_space{\baselineskip 19pt plus 1pt minus 1pt}
\def\double_spacesp{\baselineskip 25pt plus 2pt minus 2pt}

\def\atversim#1#2{\lower0.7ex\vbox{\baselineskip\zatskip\lineskip\zatskip
  \lineskiplimit 0pt\ialign{$\matth#1\hfil##\hfil$\crcr#2\crcr\sim\crcr}}}

\begin{document}
\begin{titlepage}
\begin{flushright}
{\bf
PSU/TH/185\\
July 1997\\
}
\end{flushright}
\vskip 1.5cm
{\Large
{\bf
\begin{center}
Leptoquark production in ultrahigh-energy neutrino interactions revisited
\end{center}
}}
\vskip 1.0cm
\begin{center}
M.~A.~Doncheski \\
Department of Physics \\
The Pennsylvania State University\\
Mont Alto, PA 17237  USA \\
\vskip 0.1cm
and \\
\vskip 0.1cm
R.~W.~Robinett \\
Department of Physics\\
The Pennsylvania State University\\
University Park, PA 16802 USA\\
\end{center}
\vskip 1.0cm
\begin{abstract}
 
The prospects for producing leptoquarks (LQs) in ultrahigh-energy (UHE)
neutrino nucleon collisions are re-examined in the light of recent
interpretations of HERA data in terms of leptoquark production.  
We update predictions for cross-sections for the production of first-
and second-generation leptoquarks  in UHE 
$\nu-N$ and $\overline{\nu}-N$ collisions including 
(i) recent experimental limits on masses and couplings from the 
LEP and TEVATRON colliders as well as rare processes, 
(ii) modern parton distributions, and (iii) radiative corrections
to single leptoquark production.   
If the HERA events are due to an $SU(2)$ doublet 
leptoquark which couples mainly to $e^{+}q$ states, we argue that 
there are likely other LQ states which couple to neutrinos which are
close in mass, due to constraints from precision electroweak measurements.
\end{abstract}
\end{titlepage}
\double_spacesp

The recent observation by two groups at HERA \cite{HERA} 
of an excess of  events
in deep inelastic neutral-current scattering at
large $Q^2$ in the $e^{+}p$ channel has
spawned a large number of papers which attempt to interpret the deviation
from the standard model prediction in terms of new particles.  While
many interpretations have been put forth, the possibility that
leptoquarks (LQs) of mass roughly $200\,GeV$ 
are responsible has received the most attention.  (For
recent reviews of these general ideas, 
with references to many of the original papers, see
Refs.~\cite{blumlein}, \cite{frampton},\cite{ks}, and \cite{rizzo}.)

While the production of leptoquarks in $eq$ collisions are HERA
depends on the unknown LQ-q-e coupling, pair production of leptoquarks
in $e^{+}e^{-}$ collisions  at the $Z^{0}$ pole at LEP and in hadron
collisions, which rely solely on the LQ electroweak and strong couplings
(which are restricted in form), have now set stringent mass limits.
LEP experiments \cite{LEP} have excluded leptoquarks with masses below
$\sim 45\,GeV$, independent of their couplings to leptons and quarks, 
while recent analyses of TEVATRON collider data have yielded limits
on first-generation scalar leptoquarks of $M(LQ_1) > 210\,GeV$
\cite{cdf_1}  (assuming a branching ratio to $eq$ final 
states of $\beta = 1.0$) and $M(LQ_1) \geq 147\,GeV\, (71\,GeV)$
\cite{cdf_2} 
for $\beta = 0.5\,(0.0)$ where the $\beta=0.0$ limit is important as it
provides a constraint on LQs which couple exclusively to $\nu q$ final
states. These limits include the results of NLO QCD calculations of
LQ pair-production cross-sections \cite{kramer} and are even more
stringent for vector leptoquarks (which have a larger 
strong-interaction cross-section at leading order) where 
one finds \cite{cdf_2} $M(LQ_1) > 298\,GeV\,(270\,GeV)$ for
$\beta = 1.0 \, (0.5)$.   

Kunszt and Stirling \cite{ks} have argued that these mass limits are likely 
sufficient to already 
exclude vector leptoquarks from consideration as candidates for the
HERA events, while constraints
from atomic parity violation force one to consider only iso-doublet
leptoquarks in the standard $SU_C(3)\times SU_2(L) \times U(1)$ 
invariant coupling classification scheme \cite{wyler}.  The two iso-doublet
leptoquarks which we will then consider 
can be written in the form \cite{rizzo}
\begin{equation}
S_2 =
\left(
\begin{array}{c}
S_2^{(5/3)} \\
S_2^{(2/3)} \\
\end{array}
\right)
\qquad 
\qquad
\qquad 
\tilde{S}_2 =
\left(
\begin{array}{c}
\tilde{S}_2^{(2/3)} \\
\tilde{S}_2^{(-1/3)} \\
\end{array}
\right)
\end{equation}
with couplings
\begin{equation}
\begin{array}{ll}
S_{2}^{(5/3)}:          & g_{2L}(e^{+}u)\; , \;   \; g_{2R}(e^{+}u) \\
S_{2}^{(2/3)}:          & g_{2L}(\overline{\nu}_eu) \; , \;  -g_{2R}(e^{+}d) \\
\tilde{S}_{2}^{(2/3)}:  & \tilde{g}_{2L}(e^{+}d) \\
\tilde{S}_{2}^{(-1/3)}: & \tilde{g}_{2L}(\overline{\nu}_e d) \\
\end{array}
\end{equation}
In addition, because of the stringent limits from the leptonic decays
of charged  pions and kaons, one assumes that only one chiral coupling is
present, namely that $g_{2L} << g_{2R}$ or $g_{2L} >> g_{2R}$.  Assuming
that there is only one leptoquark accessible in the HERA energy range, we
are left with three candidates for LQs which couple to $e^{+}q$ states
and various corresponding possibilities for LQs which interact with neutrinos.
If the leptoquark interpretation is correct and the state observed at HERA
is $S_2^{(5/3)} \longleftrightarrow e^{+}u$ (therefore with $g_{2L} >> g_{2R}$)
or $\tilde{S}_{2}^{(2/3)} \longleftrightarrow e^{+}d$ 
(with only $\tilde{g}_{2L}$),
then there will be an iso-doublet partner which will couple to neutrinos,
namely
$S_2^{(2/3)} \longleftrightarrow \overline{\nu}_{e} u$ 
(also with $g_{2L} >> g_{2R}$)
or $\tilde{S}_2^{(-1/3)} \longleftrightarrow \overline{\nu}_{e} d$ (with only
$\tilde{g}_{2L}$.)  Only  in the case where the state nominally seen at 
HERA is $S_{2}^{(5/3),(2/3)} \longleftrightarrow 
(e^{+}u), (e^{+}d)$ (with $g_{2R} >> g_{2L}$) will there be no state which
couples directly to $\nu_{e}-q$. 

In the first two cases above, the iso-doublet partners of the particles
discussed in the context of HERA could be produced in ultra-high energy
$E_{\nu} > M(LQ)^2/2m_N$ neutrino-nucleon collisions and this topic was
discussed some time ago by one of the authors \cite{robinett}.  In
this note, we update the discussion of Ref.~\cite{robinett} to discuss
the prospects for producing first-generation leptoquarks in 
$\nu_e, \overline{\nu}_{e}-N$ collisions, motivated by the new HERA
interpretations.  We extend the discussion of Ref.~\cite{robinett}
as well to include (i) modern parton distributions which 
include new information
on the low-$x$ parton content of the proton 
as obtained from more recent fits including
HERA data, (ii) the next-to-leading order single leptoquark production
cross-section formulae in Ref.~\cite{ks}, (iii) 
recent TEVATRON collider limits on LQ masses, 
and, most importantly, (iv) the limits
on mass splittings between the iso-doublet leptoquark states which couple
to $eq$ and $\nu q$ states which arise from precision electroweak measurements.

We will consider two scenarios in which a LQ coupled to neutrinos might
be produced in a way which is consistent with existing collider mass
bounds, limits from rare processes \cite{rare}, and precision electroweak
measurements.  Let us first assume that the HERA events are due to
a $\sim 200\,GeV$ $\tilde{S}_{2}^{2/3}$ with a required coupling
\cite{ks} of $\tilde{g}_{2L}^2 = 0.002$ or $\tilde{g}_{2L} \approx
0.045$.  The iso-doublet partner of this particle, to leading order, should
be degenerate with it, but mass splittings are possible provided they are
consistent with limits from precision electroweak measurements such as
the $\rho$ parameter.  In this case, for example, we have an almost
degenerate scalar doublet which would give a contribution 
\cite{barbieri},\cite{pdg} to the $\rho$ parameter of 
\begin{equation}
\Delta \rho = \frac{1}{2}
\left[\frac{3G_F}{8\sqrt{2}\pi^2}
\Delta m^2\right]
\end{equation}
where
\begin{equation}
\Delta m^2 = m_1^2 + m_2^2 - \frac{2m_1^2m_2^2}{m_1^2 - m_2^2}
\log\left(\frac{m_1^2}{m_2^2}\right)
\end{equation}
(This contribution is one-half that of a chiral quark doublet or that
from both chiral ($L$ and $R$) components of a squark doublet
\cite{barbieri}.)
To be conservative, if we assume that three such LQ splittings (one per
generation) are allowed, 
using fits discussed in the most recent Review of Particle Properties
\cite{langacker}, we find the bounds
\begin{displaymath}
\Delta m^2 \leq (62\,GeV)^2, (80\,GeV)^2, (100\,GeV)^2
\end{displaymath}
corresponding to a mass of a standard model Higgs boson of $M_H=80\,GeV$,
$300\,GeV$, and $1000\,GeV$ respectively.  Using the middle value
of $M_H$, we find that this implies that the
component of $\tilde{S}$ which couples to neutrinos could only be as light
as $M(\tilde{S}_2^{(-1/3)}) \approx  130\,GeV$.  This limit is already
more stringent than that set by direct searches for $\nu \nu jj$ final states
\cite{cdf_2}
due to a LQ which couples exclusively to $\nu q$ states. 
(During the completion of this
project, we became aware of Ref.~\cite{ma} which  uses more electroweak
data and the parameters  $S$ and $T$ and derives 
slightly more stringent limits on such splittings than this.  On the
other hand, it has been pointed out \cite{babu} that mixings between
various LQ states could significantly weaken such bounds, even giving rise
to negative contributions to $\Delta \rho$.)

Thus, for this scenario, we use the parameters
\begin{equation}
\mbox{Scenario I:} \qquad M(LQ) = 130\,GeV \qquad \tilde{g}_{2L} =
0.045
\label{scenario_1}
\end{equation}
We note that if the HERA events are due to the process
$e^{+} u \rightarrow S_2^{(5/3)} \rightarrow e^{+}u$ with $g_{2L} \neq 0$
(and hence $g_{2R} \approx 0.0$), a similar scenario would be possible.
However, in that case the required coupling \cite{ks} 
of the $S_2^{(5/3)}$ is much 
smaller  (because it couples to a valence $u$-quark) so that one would
use $g_{2L} \approx \sqrt{0.00049} \approx 0.022$ with resulting
cross-sections in $\nu-N$ interactions which would be $4$ times smaller.

A second scenario assumes that  the LQ which couples to neutrinos
is {\bf not} the iso-doublet partner of the one putatively seen at HERA,
but rather of one which has a mass just outside of the HERA range (say,
$M(LQ) \geq 250\,GeV$).  In this case, the $\rho$-parameter splitting limits
imply that the neutrino-coupled LQ could be only as light as
$M(LQ) \approx 180\,GeV$, but the limits on the common coupling, either
$g_{2L}$ or $\tilde{g}_{2L}$,  are now 
only constrained by rare processes.  From
the compilation of Davidson {\it et al.} \cite{rare}, 
we find that this coupling could
be as large as $g_{2L}, \tilde{g}_{2L} \approx 0.11$.  Thus, in this scenario
we use
\begin{equation}
\mbox{Scenario II:} \qquad M(LQ) = 180\,GeV \qquad
\tilde{g}_{2L} = 0.11
\label{scenario_2}
\end{equation}

We then use the standard, narrow-width approximation for the resonant
production of such LQ states in both $\nu_e-N$ and $\overline{\nu}_e-N$
collisions, namely
\begin{equation}
\sigma_{\nu N} = \frac{\pi g^2}{4M^2}[xq(x,Q^2)]
\end{equation}
where $q(x,Q^2)$ is the relevant parton distribution, evaluated at
$x = M^2/s$ and using $Q^2=M^2$.  We also include the recent
calculations of the next-to-leading order (NLO) contributions to LQ
production in $lq$ interactions  in Ref.~\cite{ks}.
We use the parton distributions of Ref.~\cite{parton},
both of which give very similar results: they are 
extended to smaller  values of $x$, beyond their fitted range,  using
the methods discussed in Ref.~\cite{quigg_2}. 
We plot the resulting cross-sections
for $\nu_{e}-N$ scattering (assuming  an isoscalar nucleon 
target with equal numbers
of $u$ and $d$-quarks) in Fig.~1 for both Scenarios I and II 
and the same quantities in Fig.~2, but for $\overline{\nu}_{e}-N$
scattering.  For comparison on each plot, we reproduce the recent
calculations by Gandhi {\it et al.} \cite{quigg_2} for the standard model 
charged current (CC) and neutral current (NC) neutrino and
antineutrino cross-sections at ultra-high energies.

Just as the more recent calculations of Gandhi
{\it et al.} find a larger $\nu-N$ and $\overline{\nu}-N$ cross-section
due to the increased size of the low-$x$ parton distributions, our
LQ production calculations are now 
consistently larger than those originally found
in Ref.~\cite{robinett}.  However, the effect of the NLO correction terms
are dominated here by a negative $\pi^2$ 
term multiplying the $\delta(1\!-\!z)$  term in the virtual
corrections which leads to a
$10-15\%$ decrease in the overall cross-section, compared to leading
order.   (This is in comparison to
the result near threshold where a different term \cite{ks} 
produces a small enhancement
in the single LQ production cross-section.)

We note that the hint of a resonant structure is only apparent in the
$\overline{\nu}_{e}$ cross-section (where the $\overline{\nu}_e$ interacts
with valence $u$ or $d$ quarks) with the LQ contribution equal to the
standard model neutral-current cross-section (at roughly $E_{\nu}
\approx 10^{5}\,GeV$) only in the most optimistic scenario.  The other
problem is that future UHE neutrino telescopes \cite{halzen}
(such as AMANDA, BAIKAL, and NESTOR) are being designed to detect
muon signatures from $\nu_{\mu}\, N \rightarrow \mu\, X$ interactions and not
necessarily to detect electrons or purely neutral current events.   

Motivated by this experimental constraint, we also examine the prospects
for the production of purely second-generation leptoquarks in neutrino
interactions via the processes, $\nu_{\mu} q \rightarrow LQ_2 \rightarrow
\nu_{\mu} q, \mu q'$.  Limits on purely second-generation scalar
leptoquark masses from hadron collider data are only 
slightly less stringent than those for the first generation case 
 with limits $M(LQ_2) > 167\,GeV$ \cite{cdf_2}
and $M(LQ_2) > 195\,GeV$ \cite{cdf_1} having been quoted, both for
$\beta = 1.0$.  The same limits on splittings of neutrino-coupled LQ
will then imply that the lightest second-generation LQ which could interact
via $\nu_{\mu} s \rightarrow LQ_2 \rightarrow \nu_{\mu} s$ would be roughly
$M(LQ) \approx 130\,GeV$.  If we consider only purely second-generation
leptoquarks, then the very stringent limits on admixtures of chiral couplings
from charged pion and kaon decays are avoided. Processes such as
$D_s = (c\overline{s}) \rightarrow \mu^{+} \, \overline{\nu}_{\mu}$ can be
used (as in Ref.~\cite{rare}) to derive limits which are of the form
$g_{L,R} \leq M(LQ_2)/380\,GeV$ on the purely second-generation scalar
leptoquark mass and couplings.  (Compare the similar limits in Ref.~\cite{rare}
on the $(11)(22)$ generation changing 
couplings derivable from limits on  
$D \rightarrow \mu \, \overline{e}$.) 
Such LQs also contribute to low-energy muon neutrino
and antineutrino neutral-current scattering and we find that the ratio
\begin{equation}
\frac{\sigma(\overline{\nu}_{\mu}\,N \rightarrow \overline{\nu}_{\mu} X)_{LQ}}
     {\sigma(\overline{\nu}_{\mu}\,N \rightarrow \overline{\nu}_{\mu} X)_{SM}}
\approx 0.7 \left(\frac{(g_{2L,2R}/M(LQ_2))}{(g/M_W)}\right)^4
\end{equation}
for the ratio of the LQ-induced contribution to the standard electroweak
expression.  If we insist that this contribution be no more than $1/2\%$ of
the standard model result, we find the slightly more restrictive bound
$g_{R,L} > M(LQ_2)/430\,GeV$.

We show examples of the resulting possible
contributions to muon-neutrino charged-current and neutral-current scattering
in Fig.~3 where we assume three scenarios: (i) purely NC scattering due to
a LQ with $M(LQ_2)=200\,GeV$ with $g_{2L,2R} = 0.3$, (ii) CC scattering due to
a LQ   which couples with equal strength to 
both $\nu_{\mu} q$ and $\mu  q'$ with
$M(LQ_2) = 200\,GeV$ and $g_{2L,2R} = 0.5$ and (iii) the same scenario as (ii)
but with a smaller coupling, namely $g_{2L,2R} = 0.1$.  We see once again that 
even the most optimistic scenarios are just at the limit of having an
observable effect on UHE neutrino interactions.  We have calculated
the LQ-induced cross-sections for a large number of masses and find that
the scaling law
\begin{equation}
\frac{
\sigma(\nu_{\mu}\,N \rightarrow \mu\,X)_{LQ}
}
{
\sigma(\nu_{\mu}\,N \rightarrow \mu\,X)_{SM}
}
\approx 1.8 g^2 \left(\frac{200\,GeV}{M(LQ_2)}\right)^{2.5}
\end{equation}
is a good representation over a wide range of masses for incident
neutrino energies which satisfy $E_{\nu} >> M^2(LQ_2)/2m_N$, i.e. far above
threshold.  Using the low-energy constraint on $g$ and $M(LQ)$ mentioned
above for the second-generation case, we then find the approximate bound
\begin{equation}
\frac{
\sigma(\nu_{\mu}\,N \rightarrow \mu\,X)_{LQ}
}
{
\sigma(\nu_{\mu}\,N \rightarrow \mu\,X)_{SM}
}
\leq 0.4 \left(\frac{200\,GeV}{M(LQ_2)}\right)^{0.5}
\label{basic_limit}
\end{equation}
Thus, even more stringent constraints
on LQ masses from collider experiments will likely push any LQ-mediated
effects in ultrahigh-energy neutrino interactions beyond an observable limit.

In conclusion, recent limits of leptoquark masses from hadron collider
experiments, supplemented by theoretical constraints on mass splittings
from precision electroweak measurements provide very strong lower bounds
on the masses of leptoquarks which couple only to neutrino-quark final
states, better than the direct limits \cite{cdf_2}.  Even more stringent
limits are likely from the upgraded TEVATRON with higher energies and
vastly increased statistics as well as from future LHC experiments
where limits \cite{dion} of the order of $M(LQ) = 750\,GeV \, (1000\,GeV)$ for
$\beta = 0.5 \, (\beta= 1.0)$ will likely be possible. Such increased
mass limits, coupled with even relatively weak low-energy constraints on
couplings can be combined, as in Eqn.~(\ref{basic_limit}),  to 
very tightly constrain any new
contribution to ultrahigh-energy neutrino interactions from leptoquark
production.

\begin{center}
{\Large
{\bf Acknowledgments
}}
\end{center}

We thank C.~Quigg for inspiring our renewed interest in this topic
and T.~Rizzo for communications about precision electroweak limits and
leptoquarks. 
One of us (M.A.D) acknowledges the support of Penn State University 
through a Research Development Grant (RDG).

\newpage
{\Large
{\bf Figure Captions}}
\begin{itemize}
\item[Fig.\thinspace 1.] The cross-section (in $mb$) for standard model
charged  current (CC) (solid curve) and neutral current (NC) 
(dashed curve) $\nu_{e}-N$ scattering for isoscalar nucleons
versus incident neutrino energy, $E_{\nu}$.  The
other solid (dashed) curves correspond to the contribution of first-generation
leptoquarks described by Scenario I (SI, in Eqn.~(\ref{scenario_1}), solid
curve) and Scenario II  (SII, in Eqn.~(\ref{scenario_2}), dashed curve).
\item[Fig.\thinspace 2.] Same as for Fig.~1 except for antineutrino-induced
processes, $\overline{\nu}_{e}-N$. 
\item[Fig.\thinspace 3.] The cross-section (in $mb$) for standard model
charged current $\nu_{\mu}\,N\rightarrow \mu^{-}\, X$ 
(solid curve) and $\overline{\nu}_{\mu}\,N\rightarrow \mu^{+}\,X$
(dashed curve) 
interactions versus incident neutrino energy, $E_{\nu}$.  Also shown are
the contributions from purely second-generation leptoquarks.  The various
cases considered are (i) purely NC interactions via $\nu_{\mu}\,s \rightarrow
LQ_2 \rightarrow \nu_{\mu}\,s$ with $M(LQ_2) = 130\,GeV$ and $g_{2L,2R} = 0.3$, (ii) charged current
interactions via $\nu_{\mu} \,s \rightarrow LQ_2 \rightarrow \mu^{-}\,c$
with $M(LQ_2) = 200\,GeV$ and $g_{2L,2R} = 0.5$ and (iii) the same as (ii) but
with the smaller couplings $g_{2L,2R} = 0.1$.

\end{itemize}
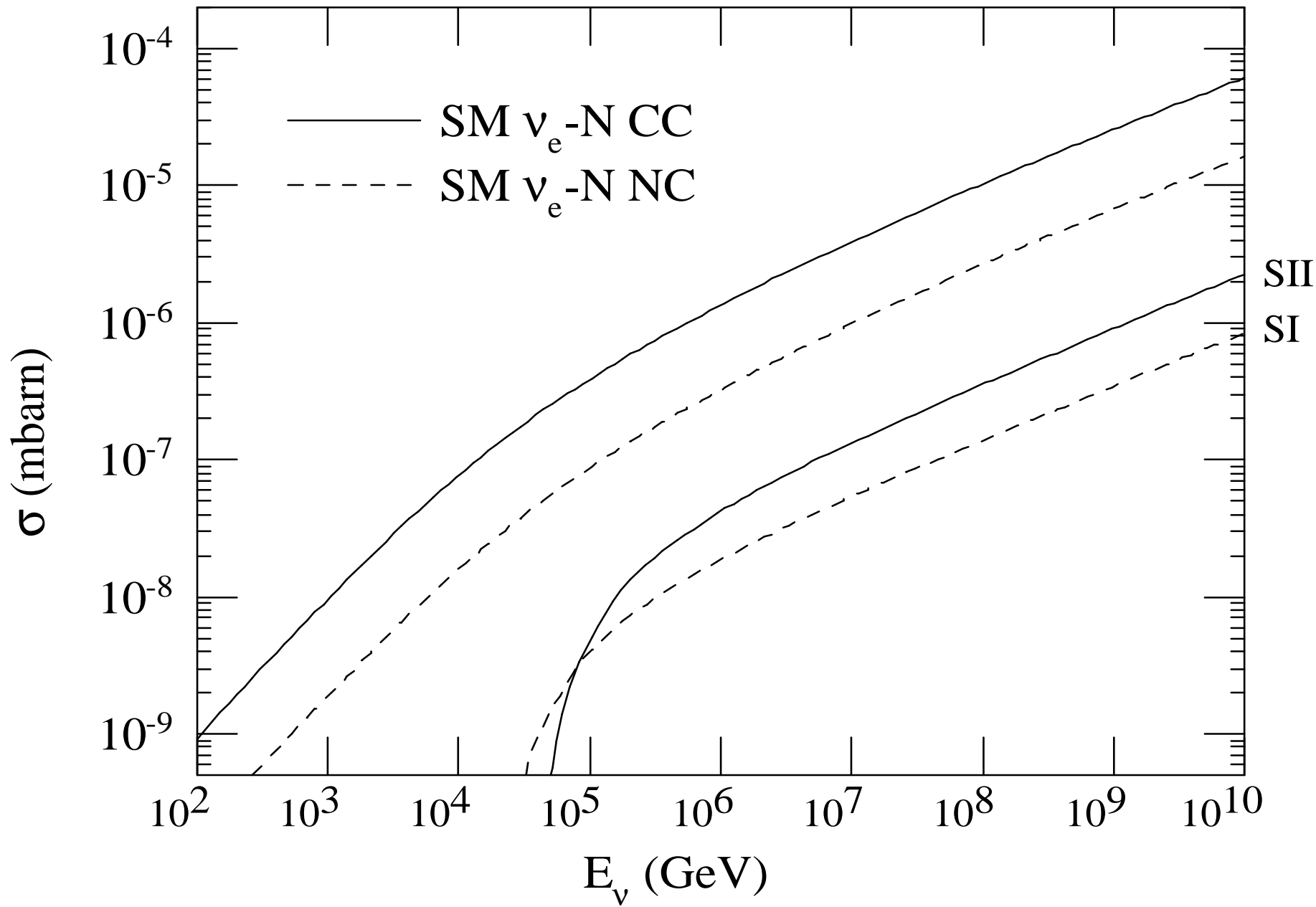
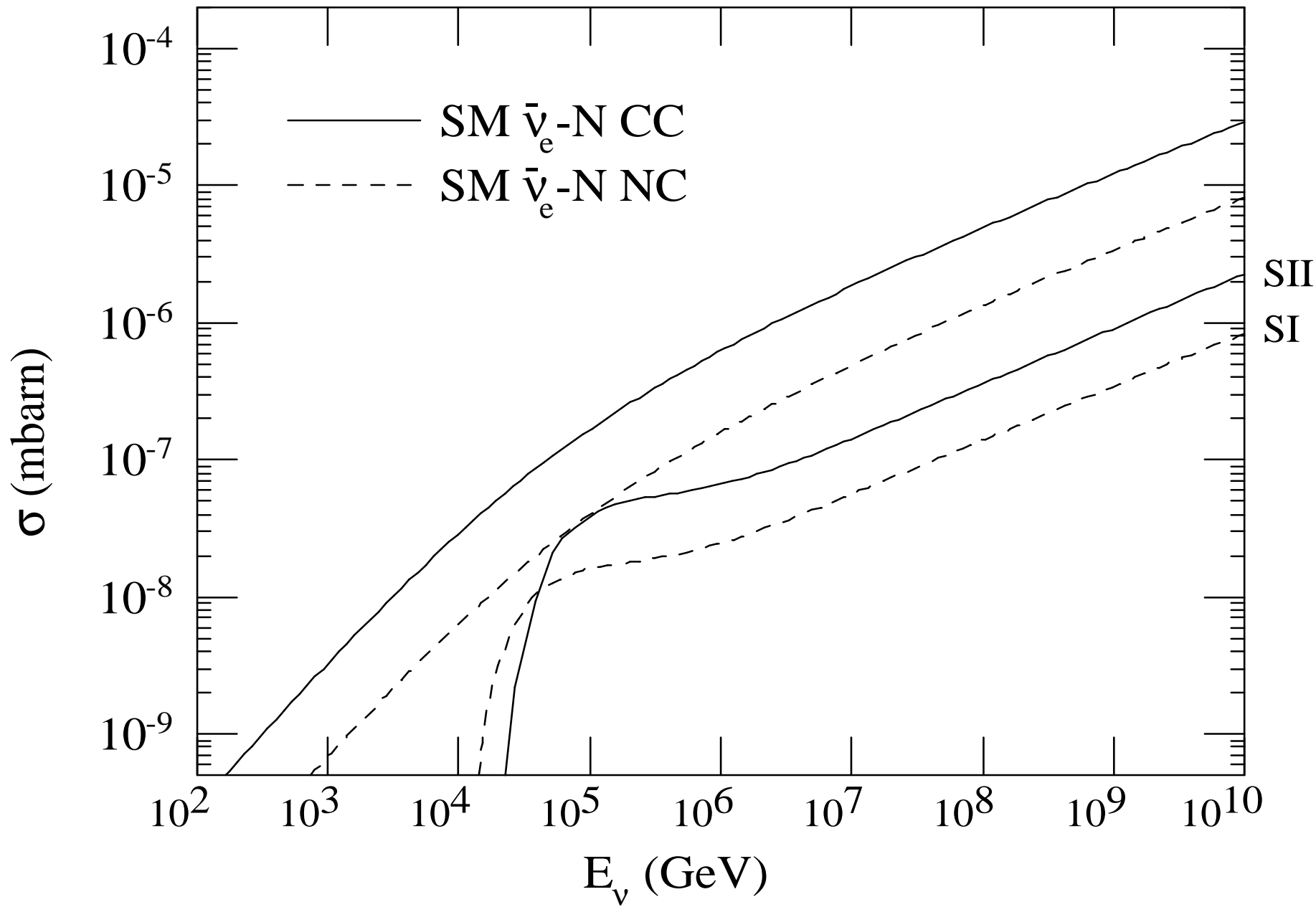
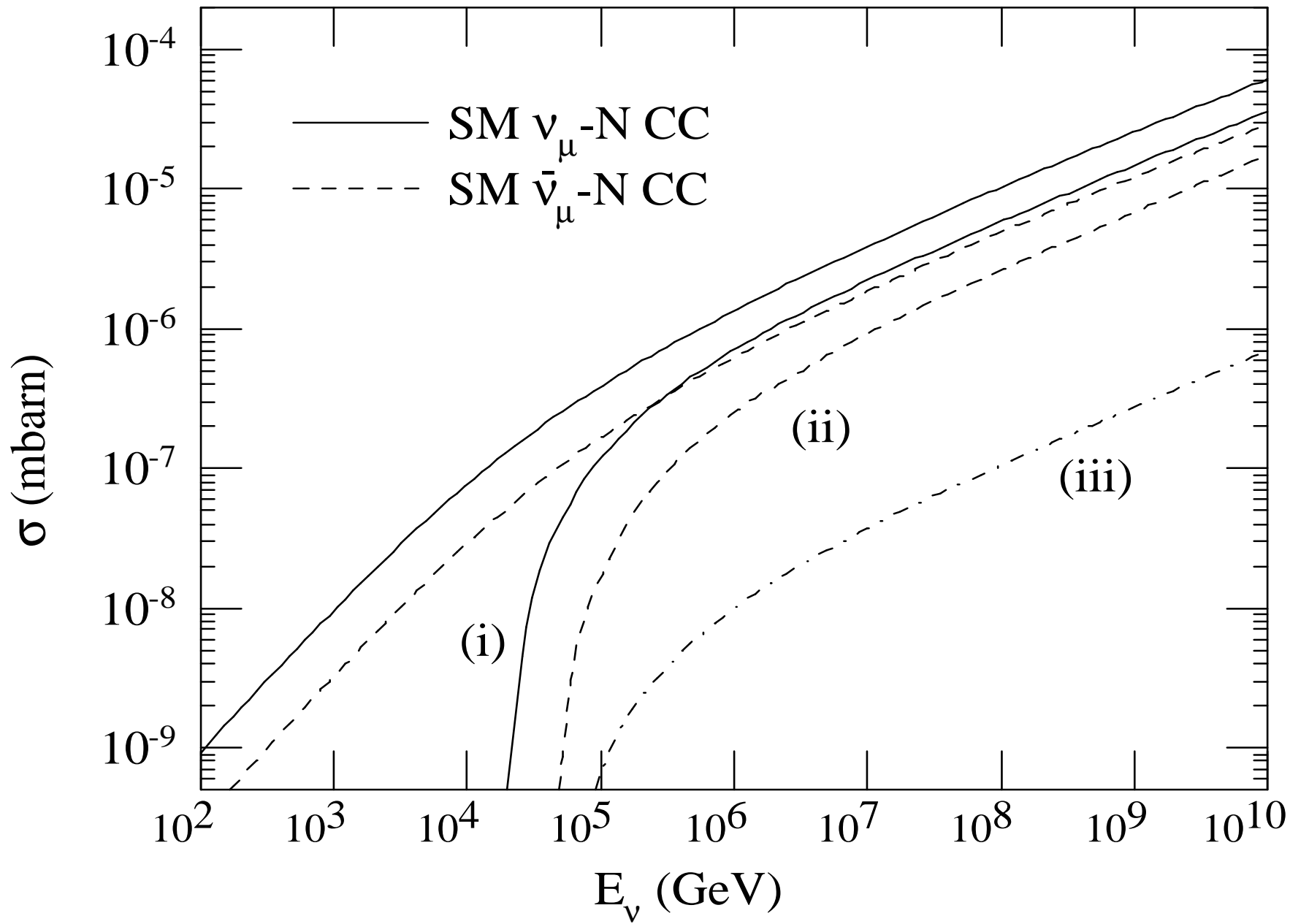
\end{document}